\documentclass{agujournal2019}
\usepackage{url} 
\usepackage{lineno}
\usepackage[inline]{trackchanges} 
\usepackage{soul}
\usepackage{graphicx}
\usepackage{wrapfig}
\usepackage{array}
\usepackage{tabularx}
\usepackage{lineno}
\usepackage{xcolor}
\draftfalse

\journalname{JGR Planets}

\begin{document}

%
%


\title{Water-Ice Dominated Spectra of Saturn's Rings and Small Moons from JWST}

\authors{M.M. Hedman\affil{1}, M.S. Tiscareno\affil{2}, M.R. Showalter\affil{2},  L.N. Fletcher\affil{3}, O.R.T. King\affil{3}, J. Harkett\affil{3}, M.T.  Roman\affil{3}, N. Rowe-Gurney\affil{4}, H.B. Hammel\affil{5}, S.N. Milam\affil{6}, M. El Moutamid\affil{7}, R.J. Cartwright\affil{2}, I. de Pater\affil{8,9}, E.M. Molter\affil{9}}

\affiliation{1}{Department of Physics, University of Idaho, Moscow ID USA}
\affiliation{2}{SETI Institute, Mountain View, CA, USA}
\affiliation{3}{School of Physics and Astronomy, University of Leicester, Leicester, UK}
\affiliation{4}{Royal Astronomical Society, UK}
\affiliation{5}{Association of Universities for Research in Astronomy, Washington, DC, USA}
\affiliation{6}{Astrochemistry Laboratory Code 691, NASA Goddard Space Flight Center, Greenbelt, MD, USA}
\affiliation{7}{Carl Sagan Institute, Cornell University, Ithaca NY USA}
\affiliation{8}{Department of Astronomy, University of California Berkeley, Berkeley CA USA}
\affiliation{9}{Department of Earth and Planetary Science, University of California Berkeley, Berkeley CA USA}

\correspondingauthor{M.M. Hedman}{mhedman@uidaho.edu}

\begin{keypoints}
\item Near-IR spectra of Saturn's small moons and rings obtained by JWST's NIRSpec show water ice bands with different degrees of crystallinity.
\item Near-IR spectra of Saturn's A ring confirm the existence of an O-D absorption band and may contain a weak aliphatic hydrocarbon band.
\item Mid-IR  spectra of Saturn's rings obtained by JWST's MIRI reveal a reflectance peak at 9.3 microns due to highly crystalline water ice.
\end{keypoints}

\begin{abstract}
JWST measured the infrared spectra of Saturn's rings and several of its small moons (Epimetheus, Pandora, Telesto and Pallene) as part of Guaranteed Time Observation program 1247. The NIRSpec instrument obtained near-infrared spectra of the small moons between 0.6 and {5.3} microns, which are all dominated by water-ice absorption bands. The shapes of the water-ice bands for these moons suggests that their surfaces contain variable mixes of crystalline and amorphous ice or variable amounts of contaminants and/or sub-micron ice grains. The near-infrared spectrum of Saturn's {A ring} has exceptionally high signal-to-noise between 2.7 and 5 microns and is dominated by features due to highly crystalline water ice. The ring spectrum also confirms that the rings possess a 2-3\% deep absorption at 4.13 microns due to deuterated water ice previously seen by the Visual and Infrared Mapping Spectrometer onboard the Cassini spacecraft. This spectrum also constrains the fundamental absorption bands of carbon dioxide and carbon monoxide and may contain evidence for a weak aliphatic hydrocarbon band. Meanwhile, the MIRI instrument obtained mid-infrared spectra of the rings {between 4.9 and 27.9 microns, where the observed signal is a combination of reflected sunlight and  thermal emission.} This region shows a strong reflectance peak centered around 9.3 microns that can be attributed to crystalline water ice. Since both the near and mid-infrared spectra are dominated by highly crystalline water ice, they should provide a useful baseline for {interpreting the} spectra of other objects in the outer solar system with more complex compositions.

\end{abstract}

\section*{Plain Language Summary}
Saturn's rings and small moons are all composed primarily of very pure water ice, making them useful targets for characterizing the performance of the various instruments onboard JWST. Observations of multiple small moons at near-infrared wavelengths demonstrate the ability of JWST to detect faint objects in the outer solar system, and reveal that the water ice on these bodies is not always organized into large, pure crystals. Observations of Saturn's rings, by contrast, confirm that they are composed of very pure and highly crystalline water ice. These data also provide new constraints on the amounts of carbon-containing compounds that could be present in the rings.

\section{Introduction}

The infrared spectra of Saturn's rings and small moons have been well characterized thanks to extensive observations obtained by the Visual and Infrared Mapping Spectrometer (VIMS) and Composite Infrared Spectrometer (CIRS) onboard the Cassini spacecraft \cite{Brown04, Flasar04}. Observations of these objects with the instruments onboard JWST therefore provide opportunities not only to obtain new information about the rings and small moons themselves, but also to evaluate the performance of JWST's instruments, particularly the Integral Field Unit (IFU) components of NIRSpec and MIRI \cite{Gardner23, Jakobsen22, Boker22, Boker23, Wright23}. 

A wide range of spectroscopic observations demonstrate that Saturn's main rings are composed of very pure and highly crystalline water ice. While there is evidence that the rings do contain variable amounts of non-icy materials that influence both the overall brightness of the rings and the shape of their spectra at visible and ultraviolet wavelengths \cite{Clark08, Clark19, Cuzzi09, Cuzzi18, Filacchione12, Filacchione14, Hedman13, Nicholson08}, detailed modeling of infrared and radio data indicate that most of the rings are over 99\% water ice \cite{Ciarniello19, Zhang19}.  {Indeed, while some observations by Cassini-VIMS have indicated that there could be weak ($\sim 3\%$) hydrocarbon features between 3.4 $\mu$m and 3.6 $\mu$m throughout the rings \cite{Filacchione14}, extremely high signal-to-noise spectra of the B ring contained no non-water-ice spectral features between 1 and 5 $\mu$m at the $\sim$1\% level, but did exhibit a 2-3\% brightness dip around 4.13 $\mu$m that could be due to deuterated water ice \cite{Clark19}.} Saturn's rings therefore provide a useful baseline for evaluating the quality and calibration of near-infrared spectra obtained by NIRSpec. 

Meanwhile, Cassini-CIRS observed the rings at wavelengths between 7.2 $\mu$m and 1 mm, and data from its longer-wavelength channels have been used to constrain the temperature of the ring material \cite{Spilker06, Altobelli08, Flandes10, Filacchione14, Spilker18}. These temperature estimates can be compared with those extracted from the MIRI observations in order to assess the current calibration pipeline. At the same time, no one has yet used the CIRS data to publish a continuous brightness spectrum of the rings between 7 and 20 $\mu$m (but see Morishima et al., 2012). The JWST MIRI data can therefore provide the first detailed information about the ring's spectral properties over the wavelength range where the signal transitions from being predominantly reflected sunlight to mainly thermal emission from the rings themselves.
\nocite{Morishima12}

In addition, Cassini-VIMS observations of the small moons revealed that their near-infrared spectra are also dominated by water ice features \cite{Filacchione10, Filacchione12, Buratti19}. Furthermore, these moons span a wide range of sizes, with the average effective radii of the representative small moons Epimetheus, Pandora, Telesto and Pallene being 58.6 km, 40 km, 12.3 km and 2.23 km, respectively \cite{Thomas20}. While all these objects are point sources to JWST, the signals from these four moons span a factor of nearly 700 because the observed flux scales with cross-sectional area. The spectra derived from these moons can therefore provide a baseline for evaluating the signal-to-noise for future observations of small bodies in the outer solar system. In addition,  NIRSpec does not have the same gaps in wavelength coverage as VIMS around 1.6 $\mu$m and 3.0 $\mu$m \cite{Brown04}, so these spectra can provide new information about the crystallinity of the ice on these objects.

For all of these reasons,  as part of JWST Guaranteed Time Observation 1247,  near-infrared spectra of Saturn's rings and a representative sample of its small moons (Epimetheus, Pandora, Telesto and Pallene) were obtained with the NIRSpec IFU, and mid-infrared spectra of Saturn's rings were obtained with the MIRI IFU.  This paper provides an overview of these observations and summarizes the early findings derived from them.  Section~\ref{nirspec} describes the NIRSpec observations of Saturn's rings and moons. This section begins with a general description of the relevant observations, which is followed by descriptions of how we extracted spectra of the small moons and rings from these data and discussions of some interesting aspects of those spectra.  Section~\ref{miri} describes the MIRI observations of Saturn's rings. This section also begins by describing the relevant observations and how they were processed to obtain mid-infrared spectra of the B ring, followed by a brief discussion of the most notable features of these spectra. Finally, Section~\ref{summary} summarizes the key findings of these preliminary analyses.

\section{Near-Infrared observations of Saturn's small moons and rings}
\label{nirspec}

\subsection{General description of observations}

JWST made NIRSpec IFU observations of the rings and small satellites using the PRISM ($R\sim100$, {$\delta \lambda \sim$ 5 nm}) mode to capture as much of the 0.6-5.3 $\mu$m spectrum as possible. These observations targeted Pandora (2022-Nov-08, 21:55-22:11UT), Epimetheus (2022-Nov-08, 22:15-22:29UT), Telesto (2022-Nov-10, 12:04-12:25UT) and Pallene (2023-Jun-20 01:14- 01:35 UT). Note the Epimetheus observations were also a repeat of an earlier, skipped observation.  Given the expected brightness of Pandora and Epimetheus, the observations of these moons were designed with 2 groups and 3 integrations in the NRSRAPID readout mode (10.7-s frames), and a 2-point nod, yielding a total exposure duration of 193.262 s.  Telesto used 15 groups and 1 integration for each dither, and Pallene used 44 groups and 1 integration with each dither. Both these observations used the NRSIRS2RAPID readout mode (14.6-s frames). This mode was intended to improve performance and sensitivity in the longer exposures, and yielded total exposure times of 466.844 s for Telesto and 1313.0 s for Pallene. The Epimetheus, Pandora and Telesto observations were processed using the JWST pipeline version 1.9.3 with CRDS context jwst\_1039.pmap, while the Pallene observations were processed using pipeline version 1.10.2 with CDS context jwst\_1094.pmap. All observations were processed using the custom calibration pipeline described in \citeA{King23}.

\begin{figure}
\resizebox{\textwidth}{!}{\includegraphics{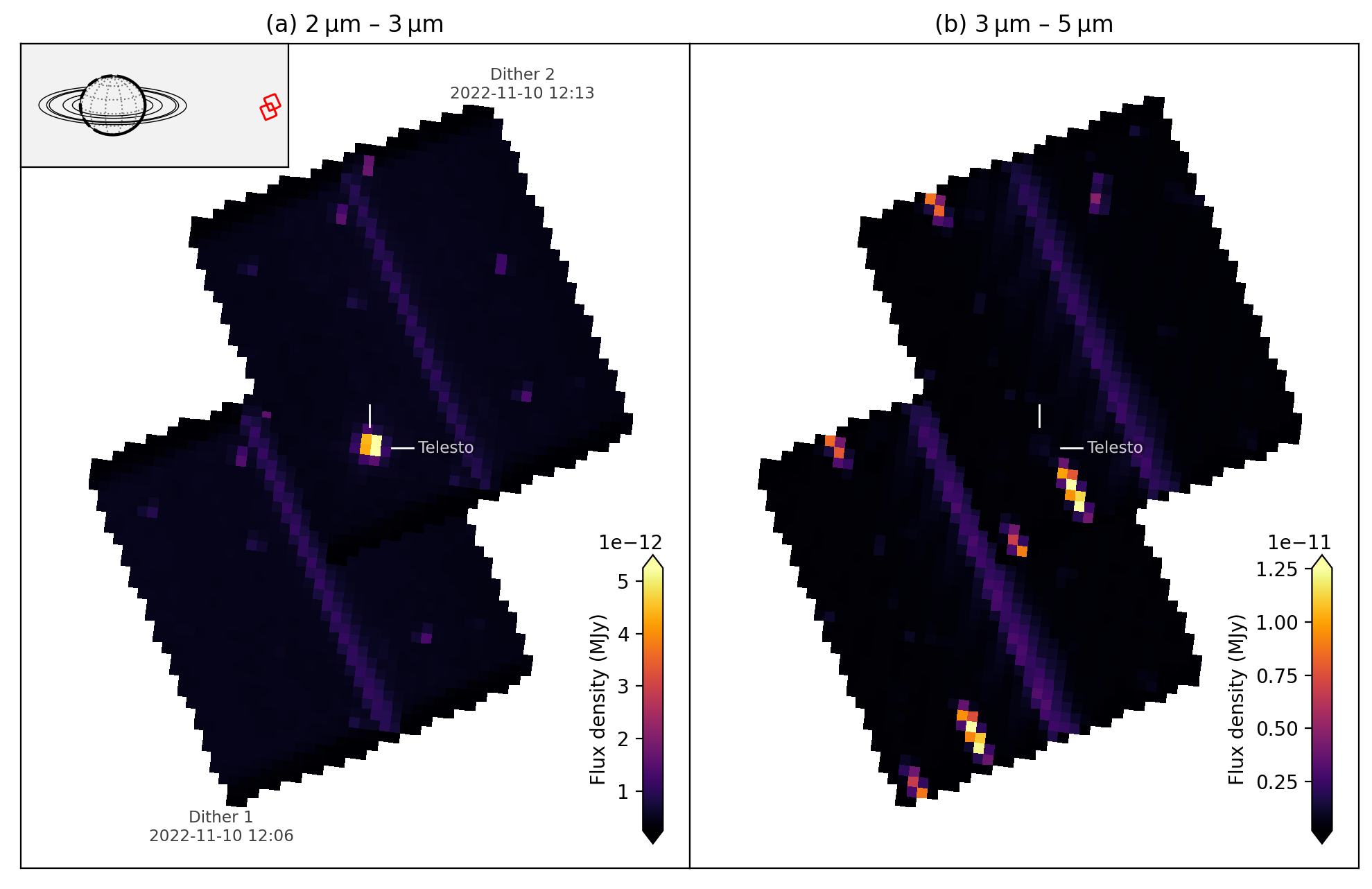}}
\caption{NIRSpec image of Telesto  {in a coordinate system centered on and aligned with Saturn.}  The inset in the upper left shows the context for the two images shown in the two panels. Each image shows the average brightness of the data derived from the two dithers over the indicated wavelength range. At 2-3 $\mu$m Telesto is clearly present, while at longer wavelengths various background instrumental artifacts are more prominent. Note that the flux density scale in these plots is per spatial pixel.}
\label{telim}
\end{figure}

\begin{figure}
\resizebox{\textwidth}{!}{\includegraphics{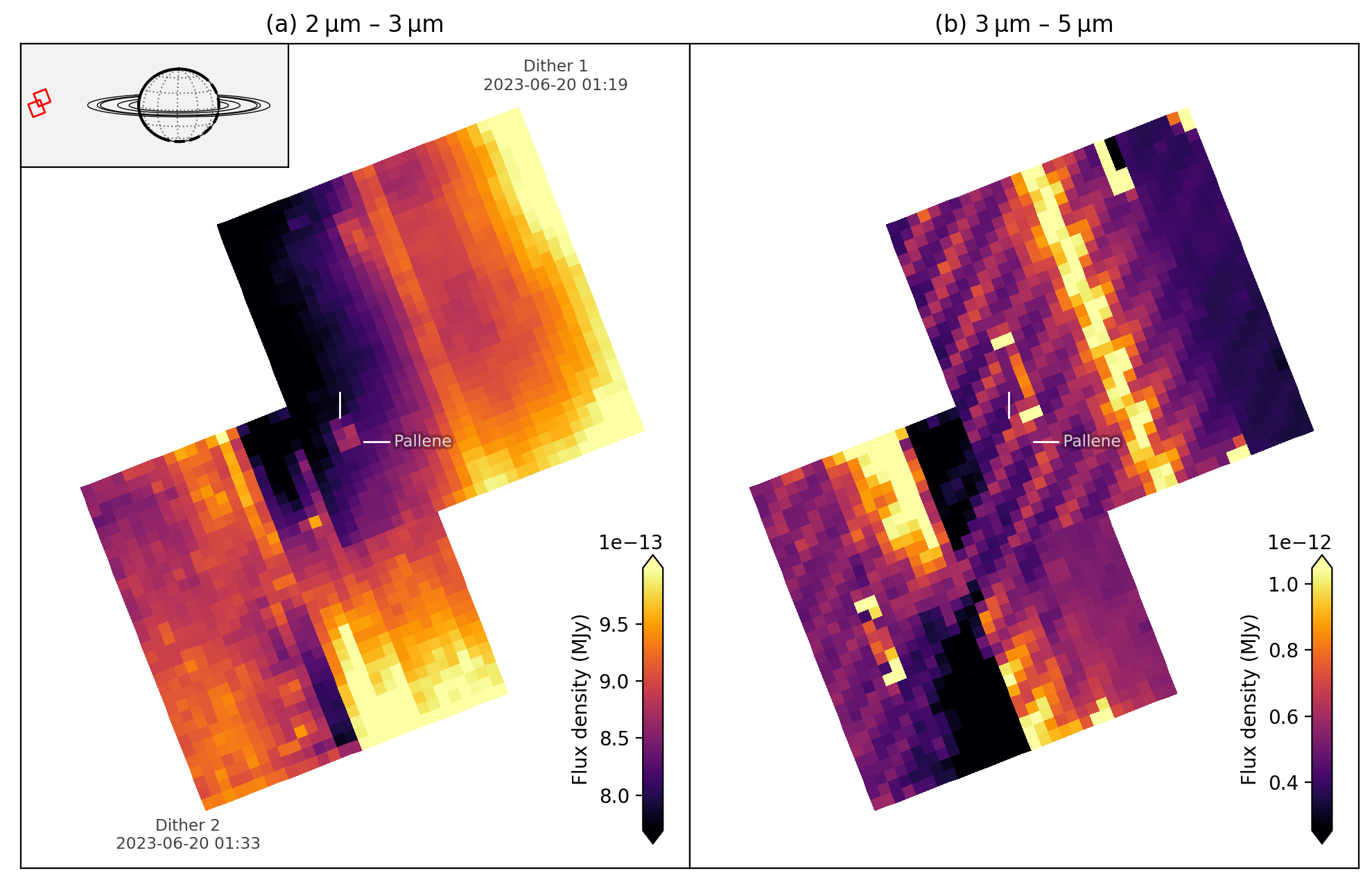}}
\caption{NIRSpec image of  Pallene  {in a coordinate system centered on and aligned with Saturn.}  The inset in the upper left shows the context for the two images shown in the two panels. Each image shows the average brightness of the data derived from the two dithers over the indicated wavelength range. At 2-3 $\mu$m Pallene is clearly present, while at longer wavelengths various background instrumental artifacts are more prominent. Note that the flux density scale in these plots is per spatial pixel.}
\label{palim}
\end{figure}

The observations targeted at Telesto and Pallene shown in Figures~\ref{telim} and~\ref{palim} are the most straightforward to interpret. Each observation consists of two dithers pointed at slightly different locations with an overlap region that should contain the moon.  For the Telesto observations, a point source is clearly visible in this overlap region  at wavelengths shorter than 3 $\mu$m (see Figure~\ref{telim}), and a fainter source is visible at nearly the same location in the Pallene observations (see Figure~\ref{palim}). The signal from Pallene is much weaker than the signal from Telesto simply because Pallene is significantly smaller (see below). Neither Pallene nor Telesto is clearly visible at wavelengths longer than 3 $\mu$m. This is consistent with both objects being composed primarily of water ice, which is a strong absorber at these longer wavelengths.  In addition to the signal from the moons, these images also show a variety of patterns that can be attributed to instrumental artifacts because they have similar spatial structures in both dithers. These patterns are more obvious in the Pallene images because the data had to be stretched further to show the signal from the smaller moon.

\begin{figure}
\resizebox{\textwidth}{!}{\includegraphics{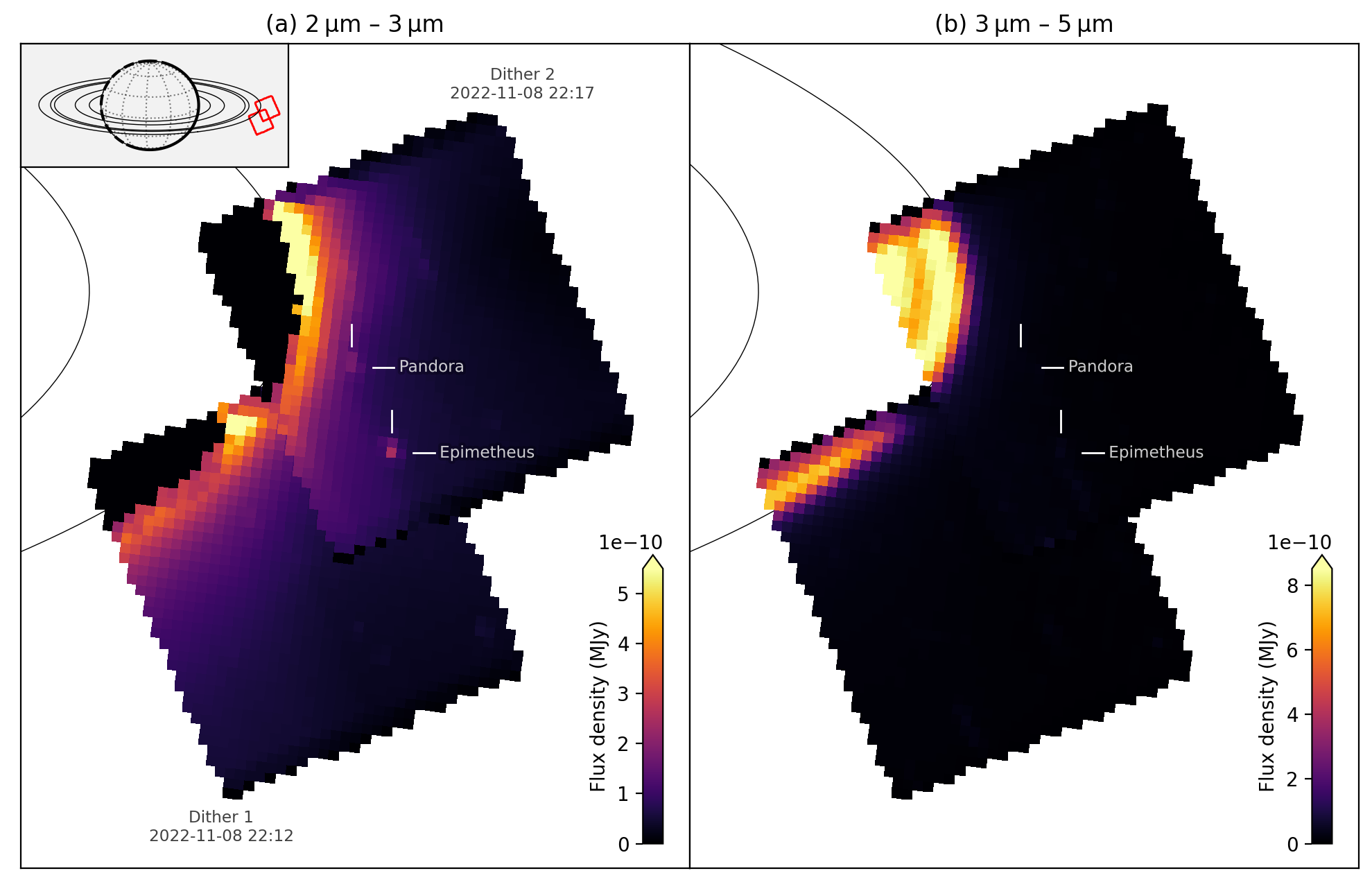}}
\caption{NIRSpec images targeted at Epimetheus {in a coordinate system centered on and aligned with Saturn.} The inset in the upper left shows the context for the two images shown in the two panels. Each image shows the average brightness of the data derived from the two dithers over the indicated wavelength range. At 2-3 $\mu$m, signals from both Epimetheus and Pandora can be identified, Saturn's A ring is also present at the top left corner of the images, which appears black in this panel because the pixels are overexposed. Above 3 $\mu$m the signals from the moons are difficult to see, but the signal from the rings are clearer because these channels are not overexposed. In particular, the Encke Gap is clearly visible. Note that the flux density scale in these plots is per spatial pixel.}
\label{epiim}
\end{figure}

\begin{figure}
\resizebox{\textwidth}{!}{\includegraphics{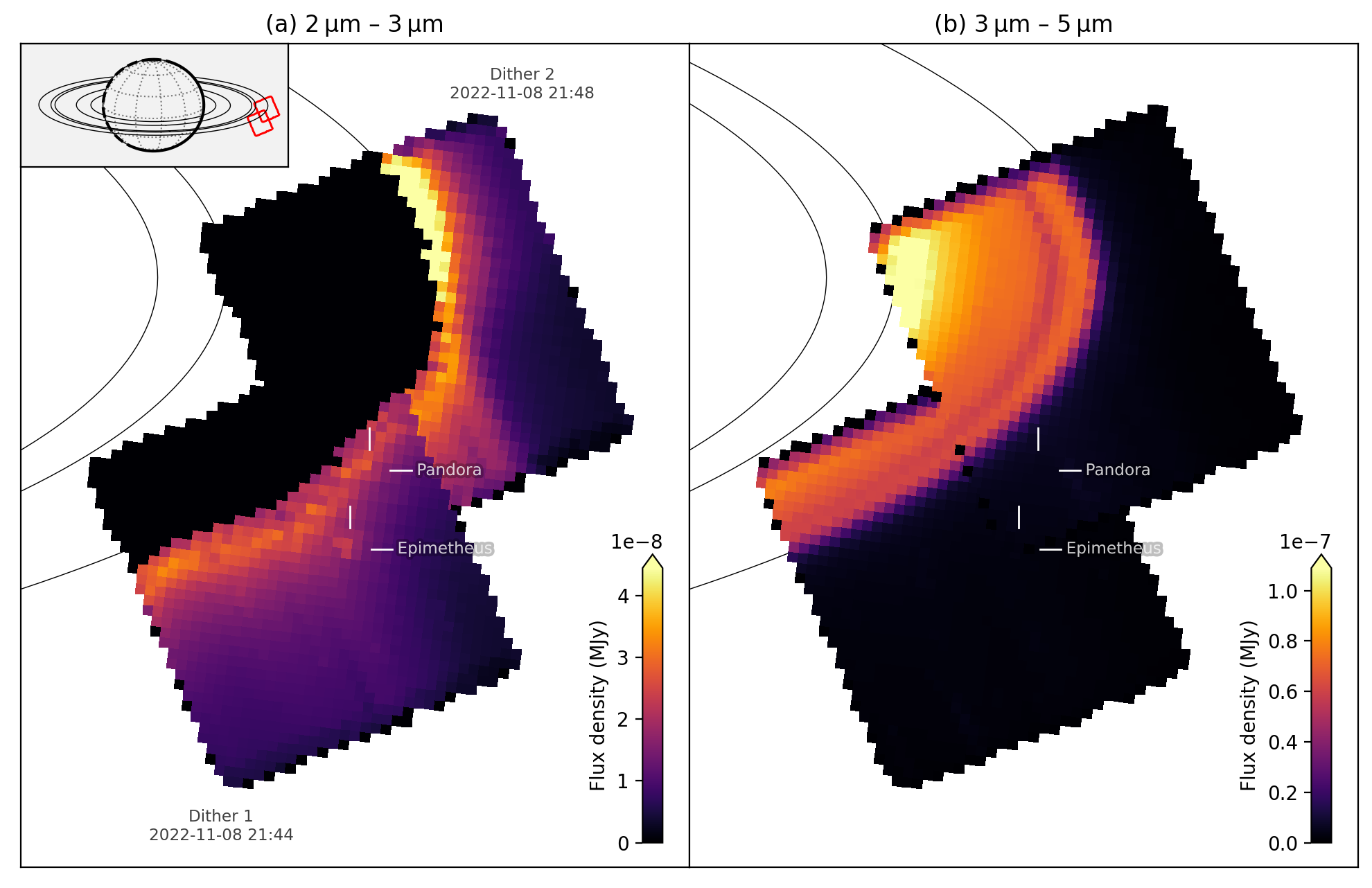}}
\caption{NIRSpec images targeted at Pandora  {in a coordinate system centered on and aligned with Saturn.} The inset in the upper left shows the context for the two images shown in the two panels. Each image shows the average brightness of the data derived from the two dithers over the indicated wavelength range. At 2-3 $\mu$m  signals from Epimetheus can be identified, and the signal from Pandora is barely visible due to contamination from the nearby ring. Saturn's A ring is also present at the top left corner of these images, but appears black in this panel because the pixels are overexposed. Above 3 $\mu$m the signals from the moons are difficult to see, but the signal from the rings is clearer because these channels are not overexposed. In particular, the Encke Gap is clearly visible in this image. Note that the flux density scale in these plots is per spatial pixel.}
\label{panim}
\end{figure}

Figure~\ref{epiim} shows the images explicitly targeted at Epimetheus, which is clearly visible in the 2-3 $\mu$m data. These images also clearly contain a portion of Saturn's rings, which are saturated at wavelengths shorter than 3 $\mu$m but are clearly visible at longer wavelengths.  In addition, a second bright spot can be seen between Epimetheus and the rings at the predicted position of the moon Pandora, so these data contain signals from both moons. Note that the regions  to the left of the field of view show a higher background brightness level than the rest of the image due to stray light from the rings.

Finally, Figure~\ref{panim} shows the images targeted at Pandora. Since these images were obtained shortly before the images shown in Figure~\ref{epiim}, the overall geometry of the two observations are similar. However, since Pandora is closer to the rings than Epimetheus, the rings take up a larger fraction of the field of view. At the same time, the stray light levels outside the rings are roughly an order of magnitude higher than they are in Figure~\ref{epiim}. This makes the moons more difficult to see in these images, with Epimetheus being just barely visible.  

\subsection{Near-infrared spectra of the small moons}

\begin{figure}
\resizebox{\textwidth}{!}{\includegraphics{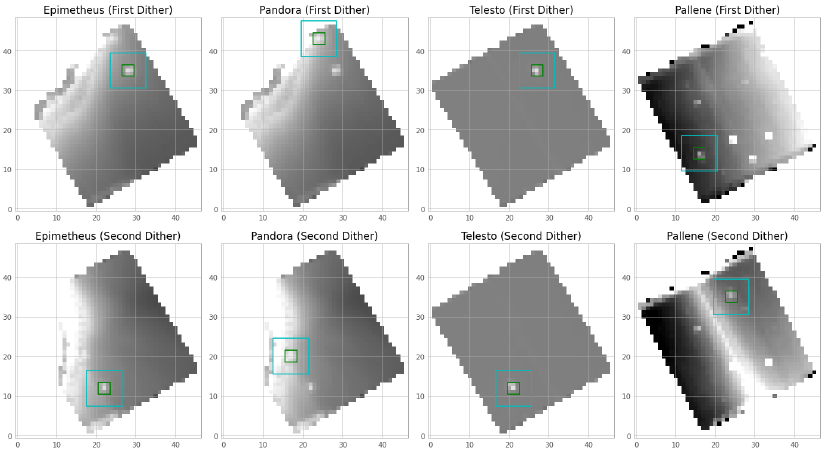}}
\caption{Images showing the regions considered for extracting the spectra of the small moons {(note the spatial dimensions of these cubes correspond to right ascension and declination, and the typical pixel size is 700 km at Saturn).} In each image, the green box indicates the 9 pixels that were co-added to obtain the spectra of the relevant moon, while a polynomial fit to the region between the cyan and green boxes was used to estimate background levels. Note the Pandora spectra were derived from the same observations as the Epimetheus spectra, and only the first dither provided a useful spectrum of Pandora.}
\label{specim}
\end{figure}

\begin{figure}
\resizebox{\textwidth}{!}{\includegraphics{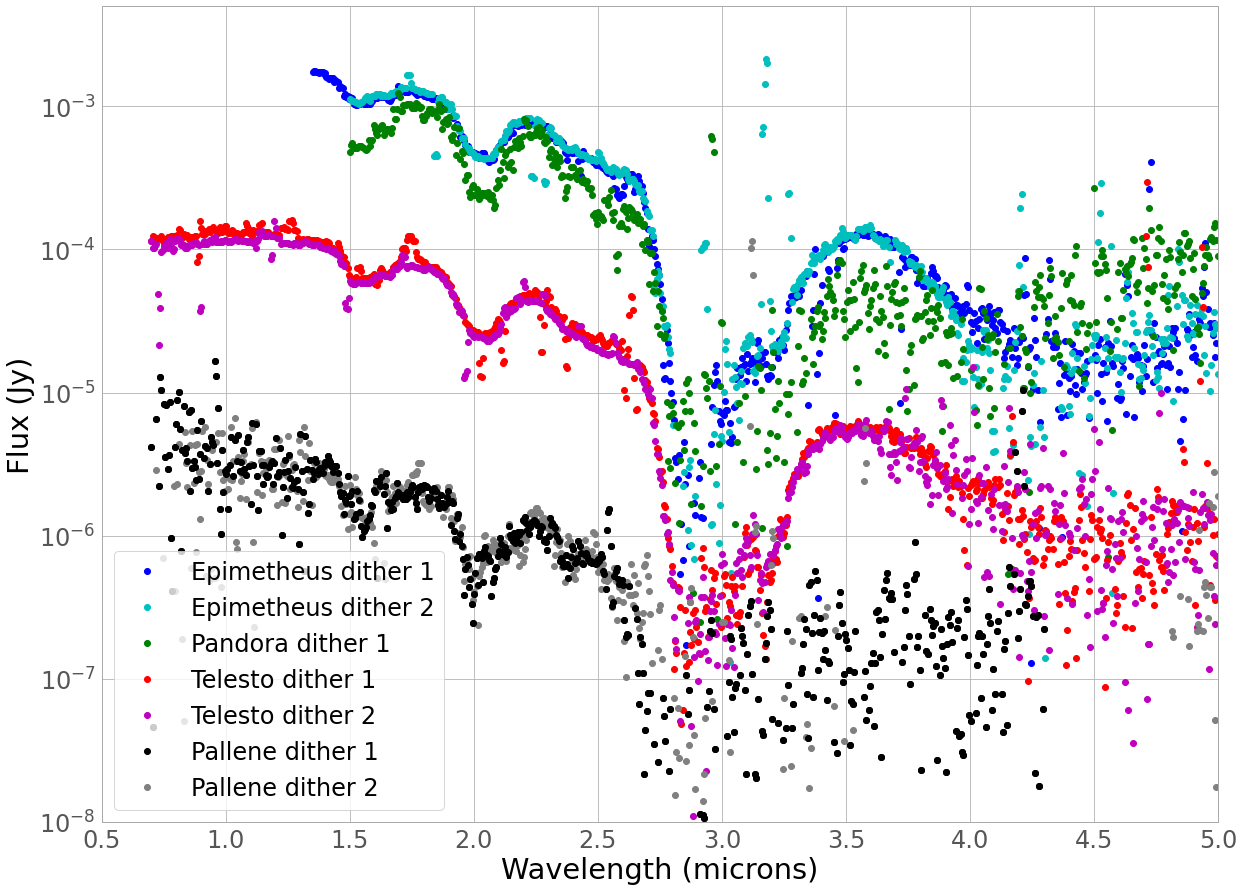}}
\caption{Flux spectra of the small moons. This plot shows the large differences in the absolute brightness of the various moons. Note that the Epimetheus and Pandora spectra end between 1.35 and 1.5 $\mu$m because the data at shorter wavelengths are contaminated by stray light from the nearby rings. Strong water-ice absorptions can be seen at 1.5 $\mu$m, 2 $\mu$m and 3 $\mu$m in all of these spectra.}
\label{fluxplot}
\end{figure}

\begin{figure}
\resizebox{\textwidth}{!}{\includegraphics{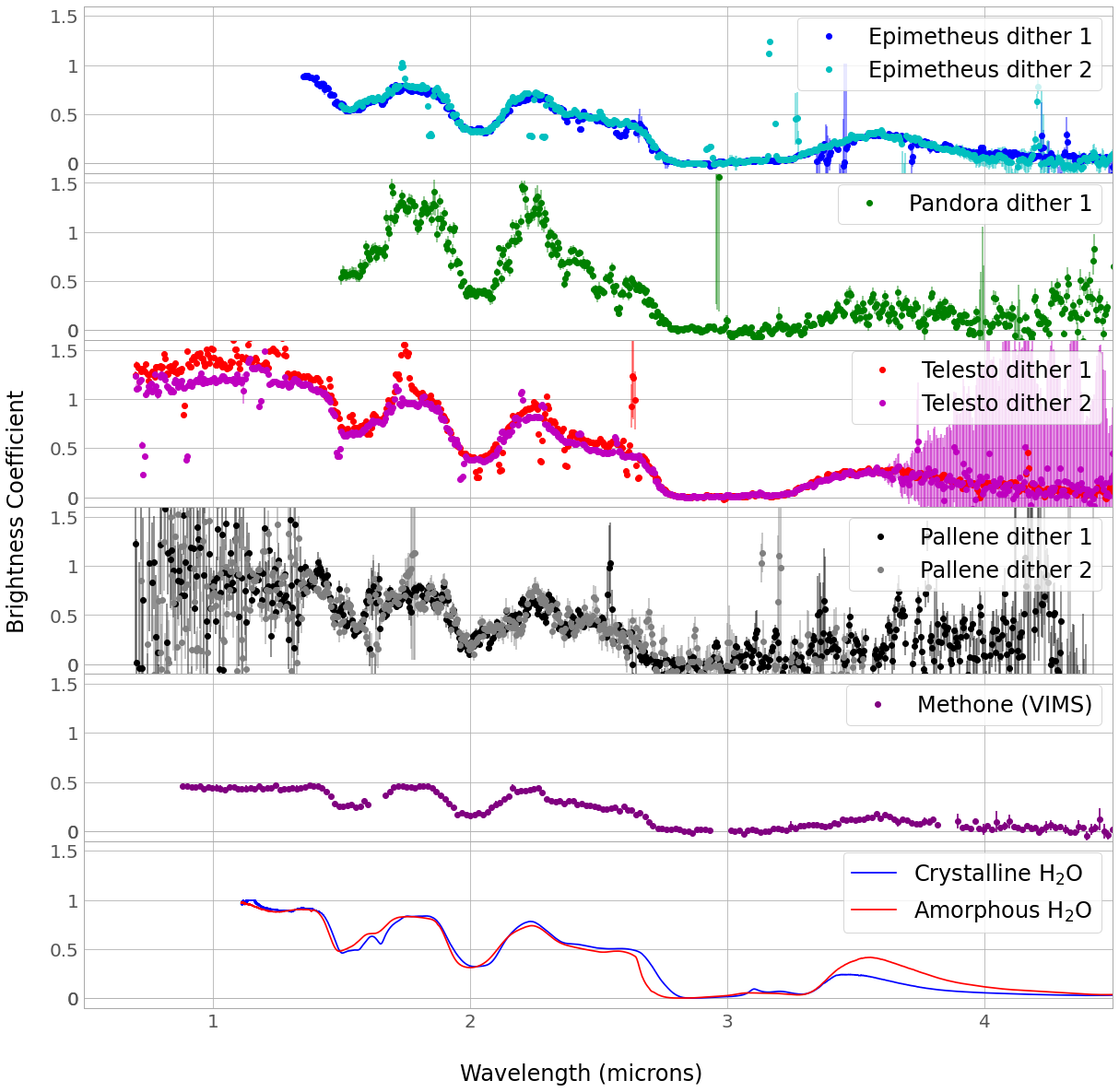}}
\caption{Near-Infrared reflectance spectra of the small moons. The spectra for each moon shows the brightness coefficient of the moon's surface computed using Equation~\ref{beq}. {Error bars show 1-$\sigma$ statistical uncertainties based on the rms brightness variations in a region surrounding each moon.} Note that the Epimetheus and Pandora spectra end between 1.35 and 1.5 $\mu$m because the data at shorter wavelengths are contaminated by stray light from the nearby rings. {For comparison, this figure also includes the reflectance spectra of Methone derived from Cassini-VIMS observations \cite{Hedman20}.} The bottom panel shows synthetic spectra for  crystalline and amorphous water ice regoliths computed using  optical constants at 100 K from \citeA{Mastrapa09} and the formulas from \citeA{Shkuratov99}, assuming a typical regolith scattering length of 10 $\mu$m.  Note the shapes and amplitudes of the 3.6 $\mu$m and 5.1$\mu$m peaks are affected by errors in the \citeA{Mastrapa09} optical constants that were identified in \citeA{Clark12}.}
\label{refplot}
\end{figure}

\begin{figure}
\resizebox{\textwidth}{!}{\includegraphics{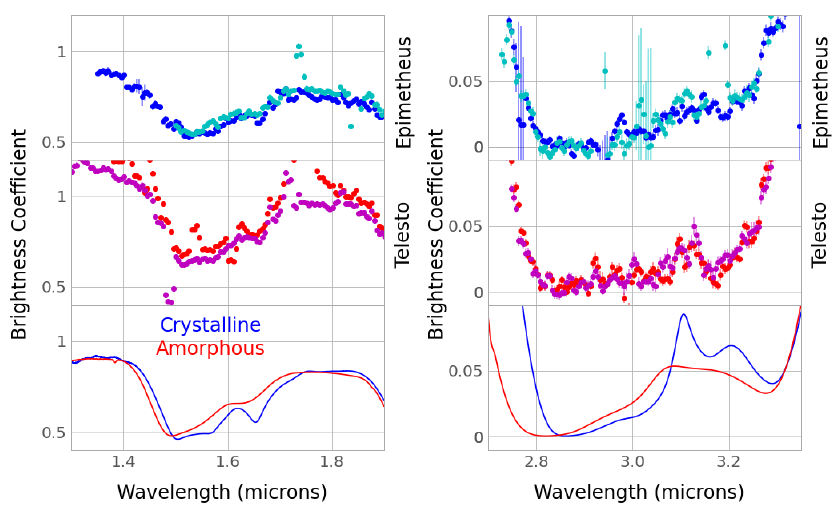}}
\caption{Close-ups of the 1.5 $\mu$m and 3 $\mu$m ice bands in the reflectance spectra of Epimetheus and Telesto, compared with the same predicted spectra for crystalline and amorphous ice regoliths shown in Figure~\ref{refplot}. Note that Telesto shows a narrower peak near 3.1 $\mu$m and a stronger dip around 1.65 $\mu$m than Epimetheus, indicating that the ice on Telesto's surface is more crystalline.}
\label{refplot2}
\end{figure}

\begin{table}
\caption{Parameters for the small-moon observations }
\label{moontab}
{\resizebox{\textwidth}{!}{\begin{tabular}{|c|c|c|c|c|c|c|c|} \hline
Parameter & Epimetheus 1 & Epimetheus 2 & Pandora 1 & Telesto1 & Telesto 2 & Pallene 1 & Pallene 2 \\
\hline
Range to Sun (10$^6$ km) & 1473 & 1473 & 1473 & 1473 & 1473 & 1465 & 1465 \\
Range to JWST (10$^6$ km) &1460 & 1460 & 1460 & 1465 & 1465 & 1402 & 1402 \\
Phase Angle (degrees) & 5.8 & 5.8 & 5.8 & 5.8 & 5.8 & 5.6 & 5.6 \\
Sub-solar latitude (degrees) & 12.6 & 12.6 & 12.7 & 12.7 & 12.7 & 9.8 & 9.8 \\
Sub-solar longitude (degrees) & 245 & 247 & 255  & 259 & 259 & 86 & 89 \\
Sub-JWST latitude (degrees) & 15.2 & 15.2 & 15.2 & 15.2 & 15.2 & 7.3 & 7.3 \\
Sub-JWST longitude (degrees) & 250 & 252 & 260 & 264 & 265 & 80 & 84 \\
$a$ (km) & 64.8 & 64.8 & 51.5 & 16.6 & 16.6 & 2.88 & 2.88 \\
$b$ (km) & 58.1 & 58.1 & 39.5 & 11.7 &11.7 & 2.08 & 2.08 \\
$c$ (km) & 53.5 & 53.5 & 31.5 & 9.6 & 9.6 & 1.84 & 1.84 \\
\hline
 Central x pixel$^*$ & 28 & 22 & 24 & 27 & 21 & 16 & 24 \\
 Central y pixel$^*$ & 35 & 12 & 43 & 35 & 12 & 14 & 35 \\
 \hline
Av. Brightness 1.36-1.40 $\mu$m & 0.876$\pm$0.011 & --- & --- &
	1.209$\pm$0.016 & 1.125$\pm$0.026 & 0.850$\pm$0.050 & 0.860$\pm$0.070 \\
Av. Brightness 1.48-1.52 $\mu$m & 0.600$\pm$0.017 & --- & --- &
	0.735$\pm$0.064 & 0.543$\pm$0.101 & 0.521$\pm$0.066 & 0.490$\pm$0.129 \\
Av. Brightness 1.77-1.81 $\mu$m & 0.741$\pm$0.009 & 0.776$\pm$0.007 & 1.180$\pm$0.112 &
				--- & 0.948$\pm$0.011 & 0.765$\pm$ 0.028 & 0.850$\pm$0.241 \\
Av. Brightness 1.84-1.88 $\mu$m & 0.714$\pm$0.018 &0.571$\pm$0.219 & 1.244$\pm$0.088 & 
	0.972$\pm$0.021 & 0.914$\pm$0.023 & 0.659$\pm$0.058 & 0.682$\pm$0.101 \\
Av. Brightness 2.03-2.07 $\mu$m & 0.336$\pm$0.017 & 0.327$\pm$0.003 & 0.386$\pm$0.013 &
	0.397$\pm$0.048 & 0.376$\pm$0.008 & 0.318$\pm$0.022 & 0.344$\pm$0.062 \\
Av Brightness 2.22-2.26 $\mu$m & 0.670$\pm$0.013 & 0.545$\pm$0.205 & 1.169$\pm$0.037 &
	0.914$\pm$0.018 & 0.817$\pm$0.005 & 0.622$\pm$0.024 & 0.724$\pm$0.056 \\
1.5 $\mu$m band depth & 0.28$\pm$0.02 & --- & --- & 
		--- & 0.49$\pm$0.09 &0.37$\pm$0.08 & 0.43$\pm$0.16 \\
2.0 $\mu$m band depth & 0.51$\pm$0.03 & 0.41$\pm$0.16 & 0.68$\pm$0.02 &
	0.58$\pm$0.05 & 0.56$\pm$ 0.01 & 0.50$\pm$0.04 & 0.51$\pm$0.10 \\
\hline
\end{tabular}}}

{$^*$ All cubes have 47$\times$49 spatial pixels.}
\end{table}

Disk-integrated spectra of the small moons were computed for each dither by co-adding the signal from the nine pixels within the green squares shown in Figure~\ref{specim} (note the central pixels are listed in Table~\ref{moontab}). after removing estimates of the instrumental background signal levels.  { The background signal level at each wavelength was derived by fitting a two-dimensional cubic polynomial function to the region between the cyan and green squares shown in Figure~\ref{specim}, excluding regions that fell outside the field of view. The rms variations of the residuals from the fit in this region are also used to estimate the statistical uncertainties of each moon's brightness at each wavelength.}  

The signal from the moon extends beyond the nine pixels used in this analysis. We did consider larger regions in the flux calculation to capture a larger fraction of the moon's signal, but found that this did not change the signal levels at short wavelengths by more than a few percent and also caused the signal levels at wavelengths above 3.5 $\mu$m to strongly diverge from the expected levels based on prior measurements \cite{Filacchione12}. We therefore elected to use a more limited number of pixels that provide more stable information about the moons' spectral shape {and estimated the missing flux using the NIRSpec observations of the solar analog star SNAP-2 (2MASS J16194609+5534178) that were obtained as part of program 1128 \cite{Gordon22}. Specifically, we computed the ratio of the stellar signal within the nine pixels centered on the star to the integrated brightness in all the observed pixels for each wavelength channel (after removing the median brightness in the entire image), and fit the resulting curve to a six-degree polynomial over the entire wavelength range, excluding outliers. The resulting function was $R =0.5183+0.5545\lambda-0.4552\lambda^2+0.1943\lambda^3-0.04713\lambda^4+0.0059\lambda^5-0.0003\lambda^6$ (where the wavelength is measured in microns) and ranges between 0.8 and 0.6 across the observed wavelength range. For this initial investigation we simply divided the observed flux from each moon by this ratio in order to approximate the moon's total flux.}

It is also worth noting that the Pandora spectrum comes from one of the dithers targeted at Epimetheus. This is because the signal from Pandora could be better isolated from the ring backgrounds in those images due to the overall background signal level being  substantially lower (compare Figures~\ref{epiim} and~\ref{panim}). Even so, we found that we could not extract a reliable spectrum of Pandora for dither 2 using the above procedure, so for the rest of this analysis we will only consider the spectrum of Pandora derived from dither 1. 

Figure~\ref{fluxplot} shows the derived flux spectra from each of the moons.  The spectra of Epimetheus and Pandora from dither 1 only extend down to 1.5 $\mu$m because shorter wavelengths were saturated by stray light from the nearby rings, while the spectrum of Epimetheus from dither 2 extends down to  1.35 $\mu$m before saturation becomes an issue. The spectra of Telesto and {especially} Pallene also become noticeably noisier at wavelengths shorter than 1.3 $\mu$m, perhaps because of residual contamination from stray light from the rings. Instrumental artifacts associated with errors in the background removal are also responsible for signal levels from certain channels being obvious outliers from the mean trend (most notably the narrow peak in the Telesto dither 1 spectrum around 1.75 $\mu$m). Note that these outliers occur in different channels for the two dithers, so they cannot be real spectral features. {These outliers also do not always correspond to wavelengths with high statistical uncertainties, indicating that these features probably involve localized systematic issues with those channels.} More careful processing could flag or remove these outliers, but for this preliminary analysis we will simply note their existence and use them to document the level of potential instrumental artifacts in these data.

Leaving aside the instrumental artifacts, the flux spectra for all the moons show clear dips around 1.5 $\mu$m, 2 $\mu$m and 3 $\mu$m consistent with the expected signatures of water ice. The different fluxes from the different moons are also reasonably consistent with their different sizes \cite{Thomas20}, with Epimetheus (average radius 58.6 km) being {somewhat} brighter than Pandora (average radius 40.0 km), and both of those moons being over an order of magnitude brighter than Telesto (average radius 12.3 km), and almost three orders of magnitude brighter than Pallene (average radius 2.23 km). 

These spectra can be more effectively compared to prior work by computing the effective average reflectance spectra of the different moons. In principle, these reflectances can be expressed in terms of the {(phase-dependent) full-disk} average $I/F$ of the moon, which is given by:

\begin{equation}
\langle I/F\rangle = \frac{\mathcal{F}_{\rm obs}}{F_\odot (1 AU)}\left(\frac{D_\odot}{1 AU}\right)^2 \frac{D_{\rm obs}^2}{\pi R_{\rm eff}^2}
\end{equation}
where $\mathcal{F}_{\rm obs}$ is the measured flux from the moon, $F_\odot(1 AU)$ is the solar flux density {(flux divided by $\pi$)} at 1 AU, $D_\odot$ is the distance from the moon to the Sun, $D_{\rm obs}$ is the distance from the moon to the observer, and $R_{\rm eff}$ is the effective average radius of the moon. However, while this is a reasonable quantity for nearly spherical moons, for elongated moons like Pandora and Pallene this quantity can vary significantly depending on how the moon is viewed due to differences in the distribution of incidence and emission angles across the visible surface \cite{Helfenstein89, Muinonen15, Hedman20}. Hence we will instead consider the average (phase-dependent) brightness coefficient of each moon $B$, which is given by the following formula:

\begin{equation}
B = \frac{\mathcal{F}_{\rm obs}}{F_\odot (1 AU)}\left(\frac{D_\odot}{1 AU}\right)^2 \frac{D_{\rm obs}^2}{a_{\rm pred}}
\label{beq}
\end{equation}
where $a_{\rm pred}$ is the predicted effective cross-sectional area of the moon computed using the python code in \citeA{Hedman20}. This code takes as input the ellipsoidal shape parameters of the moon along with the sub-solar and sub-observer latitudes and longitudes of the moon. For this particular calculation we use shape parameters from \citeA{Thomas20} and estimate the observation geometry using tools checked against the PDS Ring-Moon Systems node website (all parameters are provided in Table~\ref{moontab}). We also assume the moon's surface follows a Lambertian scattering law to facilitate comparisons with prior calculations, and we use a standard solar spectrum available from STScI (https://archive.stsci.edu/hlsps/reference-atlases/cdbs/grid/solsys/) to compute $F_\odot$ for each spectral channel. {Due to the comparatively low signal-to-noise of these satellite spectra, the solar spectrum is just evaluated at the center of each wavelength band.}

Figure~\ref{refplot} shows the derived reflectance spectra for the various moons.  Meanwhile, Table~\ref{moontab} provides the average brightness coefficients in several different wavelength regions around the 1.5 $\mu$m and 2 $\mu$m bands, with 1-$\sigma$ error bars based on the observed scatter in the data points within each wavelength. We also provide estimates of the band depths of the 1.5 $\mu$m and 2 $\mu$m bands. These band-depths are computed using the standard formula  $1-B_b/B_c$, where $B_b$ is the brightness in the middle of the relevant band and $B_c$ is the average brightness on either side of the band. For this particular study we use continuum regions consistent with prior work by \citeA{Filacchione12}.  

The overall shapes of all these satellite spectra are dominated by water-ice absorption bands at 1.5 $\mu$m, 2 $\mu$m, 3  $\mu$m and 4.5 $\mu$m. The spectra of Epimetheus, Pandora and Telesto also appear to be reasonably consistent with previously published spectra of these moons derived from Cassini-VIMS observations \cite{Filacchione12}. For Telesto, our estimates of the 1.5 $\mu$m and 2.0$ \mu$m band depths are 0.49$\pm$0.10 and 0.57$\pm$0.01, which are consistent with the \citeA{Filacchione12} values of 0.41$\pm$0.01 and 0.58$\pm$0.01. For Pandora we obtain a 2 $\mu$m band depth of 0.68$\pm$0.01, which is consistent with the \citeA{Filacchione12} value of 0.75$\pm$0.11.  For Epimetheus, our estimates of the 1.5$\mu$m and 2.0$\mu$m band depths are 0.28$\pm$0.02 and 0.51$\pm$0.03, while \citeA{Filacchione12} find values of 0.35 and 0.58 (without error bars) for Epimetheus and 0.37$\pm$0.03 and 0.58$\pm$0.02 for Janus. The ice bands on Epimetheus measured by JWST therefore appear to be slightly weaker than those observed by Cassini-VIMS. This could in part be because the JWST data were obtained at a lower phase angles ($\sim$6$^\circ$) than the VIMS observations ($>25^\circ$). Note that other moons demonstrate a reduction in ice-band depth at phase angles below 10$^\circ$ \cite{Filacchione12}.

Figure~\ref{refplot} also shows the predicted spectra of regoliths composed of pure crystalline or amorphous water ice at 100 K, computed using optical constants from \citeA{Mastrapa09} and the formulas from \citeA{Shkuratov99} that translate these constants into estimates of the surface albedo, assuming typical regolith scattering lengths of 10 $\mu$m.  Both model spectra show clear absorption bands at 1.5 $\mu$m, 2 $\mu$m and 3 $\mu$m, but there are some notable differences in the shapes of these bands. For the 1.5 $\mu$m band, crystalline ice shows a distinct secondary band on its long-wavelength flank, while amorphous ice shows a more subtle shoulder in the same region. The shape of the 2 $\mu$m band also changes, with the minimum of this band shifting to shorter wavelengths for amorphous ice. Finally, crystalline ice shows a distinct peak at 3.1$\mu$m, while amorphous ice just shows a broad hump. These features have been used to estimate the crystallinity of the icy surfaces of Saturn's larger moons \cite{DalleOre15}, {but it is worth noting that efforts to quantify the crystallinity of the ice can be complicated by diffraction from small regolith grains \cite{Clark12} and the potential presence of non-ice contaminants \cite{Ciarniello21}.}

Figure~\ref{refplot2} provides close-ups of the 1.5 $\mu$m and 3 $\mu$m bands for the higher signal-to-noise spectra of Telesto and Epimetheus, along with the reference model spectra of crystalline and amorphous water ice. Both these moons lack strong peaks at  3.1 $\mu$m and do not show a strong secondary minima on the long-wavelength side of the 1.5 $\mu$m band. This implies that the regoliths of both moons contain some amount of either amorphous or very fine-grained ice. However, these data also show that Epimetheus and Telesto have bands with different shapes. Most noticeably, Telesto's 3-$\mu$m band does contain a narrow peak near 3.1 $\mu$m, while Epimetheus' band does not, which suggests that Telesto has a larger fraction of crystalline ice than Epimetheus. The detailed shape of the 1.5 $\mu$m band is also consistent with Telesto's regolith having a higher degree of crystallinity, with a stronger 1.65$\mu$m dip and a flatter band minimum between 1.5 and 1.55 $\mu$m.  While further work is needed to properly quantify the crystallinity of these moons, these preliminary findings demonstrate that  Saturn's small moons can exhibit {crystalline signatures with a range of strengths.}

The JWST data also provides the first near-infrared spectra of the small moon Pallene. This moon's spectrum is dominated by the same water-ice features as the other moons, but the average reflectance of the moon's surface is lower than those of the other moons, which is consistent with previous photometric analyses using data at visible wavelengths \cite{Hedman20}. Furthermore, the 1.5  $\mu$m and 2 $\mu$m band depths for this moon are 0.37$\pm$0.08 and $0.50\pm0.04$, which are comparable to the band depths for the nearby and comparably small moon Methone (that is, 0.45$\pm$0.04 and 0.60$\pm$0.06, see Hedman et al. 2020). This demonstrates that JWST can provide sensible spectral data for extremely small ice-rich objects in the outer solar system. 

Furthermore, while the  Pallene spectrum  does not have sufficient signal-to-noise to document the detailed shape of the 1.5 $\mu$m and 3 $\mu$m ice bands, its 2 $\mu$m band has its minimum displaced towards shorter wavelengths than seen in the other moons observed by JWST (see Figure~\ref{refplot}). {The wavelength of Pallene's 2-$\mu$m band minimum is also similar to that seen in the spectrum of Methone previously observed by Cassini-VIMS  \cite{Hedman20} shown in Figure~\ref{refplot}. (Note that the Methone spectrum in Figure~\ref{refplot} has been re-normalized from relative brightness to brightness coefficients. The brightness coefficient of Methone is somewhat lower than for Pallene in part because the Cassini-VIMS data were obtained at higher phase angles $\sim 58^\circ$.) Since this band minimum location is also consistent with amorphous water ice, this suggests that the ice on Pallene's (and Methone's) surface may have a much higher amorphous fraction than the other small moons observed by JWST.  This finding is interesting because both Methone and Pallene experience 20-40 times higher flux of high-energy radiation than most of Saturn's moons (including Epimetheus and Telesto, cf. Hedman et al. 2020) and such radiation can cause crystalline ice to become amorphous \cite{MH92, Fama10}. Pallene's and Methone's spectra may therefore provide useful information about how efficiently radiation can influence the crystallinity of water ice in the Saturn system. {However, it is important to note that the limited signal-to-noise of these data and the relatively subtle changes in the predicted shapes of the bands mean that more careful analysis will be needed to properly quantify the significance and robustness of these variations in the band minimum position, and translate these measurements into quantitative estimates of the ice's state (cf. Dalle Ore et al. 2015).}

\subsection{Near Infrared Spectra of Saturn's A Ring}

The NIRSpec observations targeted at Pandora also provided exceptionally high signal-to-noise spectra of one ansa of Saturn's A ring {at a phase angle of 5.8$^\circ$ and a ring opening angle of 15.2$^\circ$ with a typical pixel scale of 700 km.} We computed the average ring spectrum from the second dither of the Pandora observations by taking the average of the $\sim$400 pixels with a signal higher than $1.5\times10^{-9}$ MJy between 3.5 and 3.6 microns. Again, the flux measurements are converted to reflectance using the standard solar spectrum (https://archive.stsci.edu/hlsps/reference-atlases/cdbs/grid/solsys/), yielding the spectrum shown in Figure~\ref{ringspec}. {Note that in this case we average the solar flux over the nominal wavelength band for each channel in order to better model the subtle signals from various solar lines.} We also compute the statistical uncertainty on this spectrum by computing the standard deviation of the same data points after scaling to remove overall brightness variations across the scene. Note that the rings completely saturated the detector at all wavelengths below 2.71 $\mu$m, so the available spectrum only covers longer wavelengths. The signal-to-noise of this spectrum is much higher than that of any of the small moons because the A ring fills many pixels, with the statistical uncertainties being below 0.2\% throughout most of the observed wavelength range.

\begin{figure}
\resizebox{\textwidth}{!}{\includegraphics{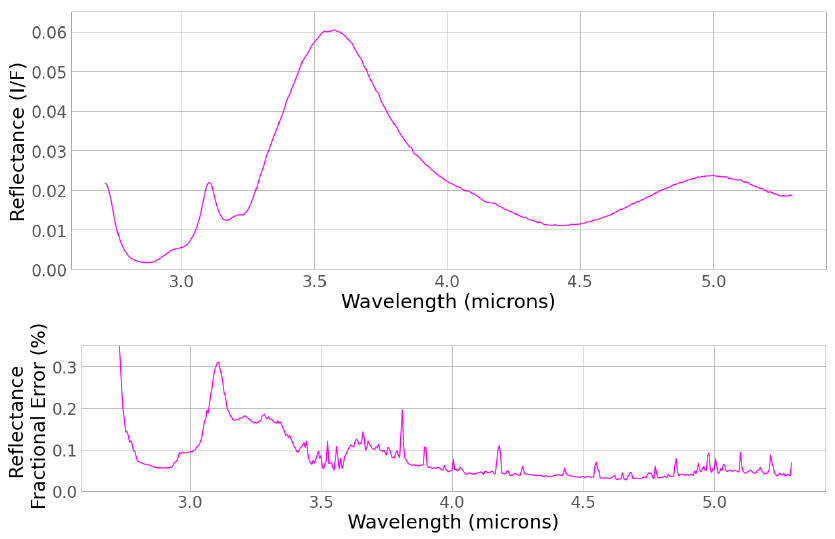}}
\caption{Reflectance spectra of the A ring obtained from the second dither of the Pandora observations. Note the spectrum only extends down to 2.71 $\mu$m because the data at shorter wavelengths were saturated. The multiple peaks around 3.1 $\mu$m are diagnostic of highly crystalline water ice (compare to Figure~\ref{refplot2}). The bottom panel shows the statistical uncertainty in the spectra based on the observed scatter in the brightness measurements among the relevant pixels.}
\label{ringspec}
\end{figure}

\begin{figure}
\resizebox{\textwidth}{!}{\includegraphics{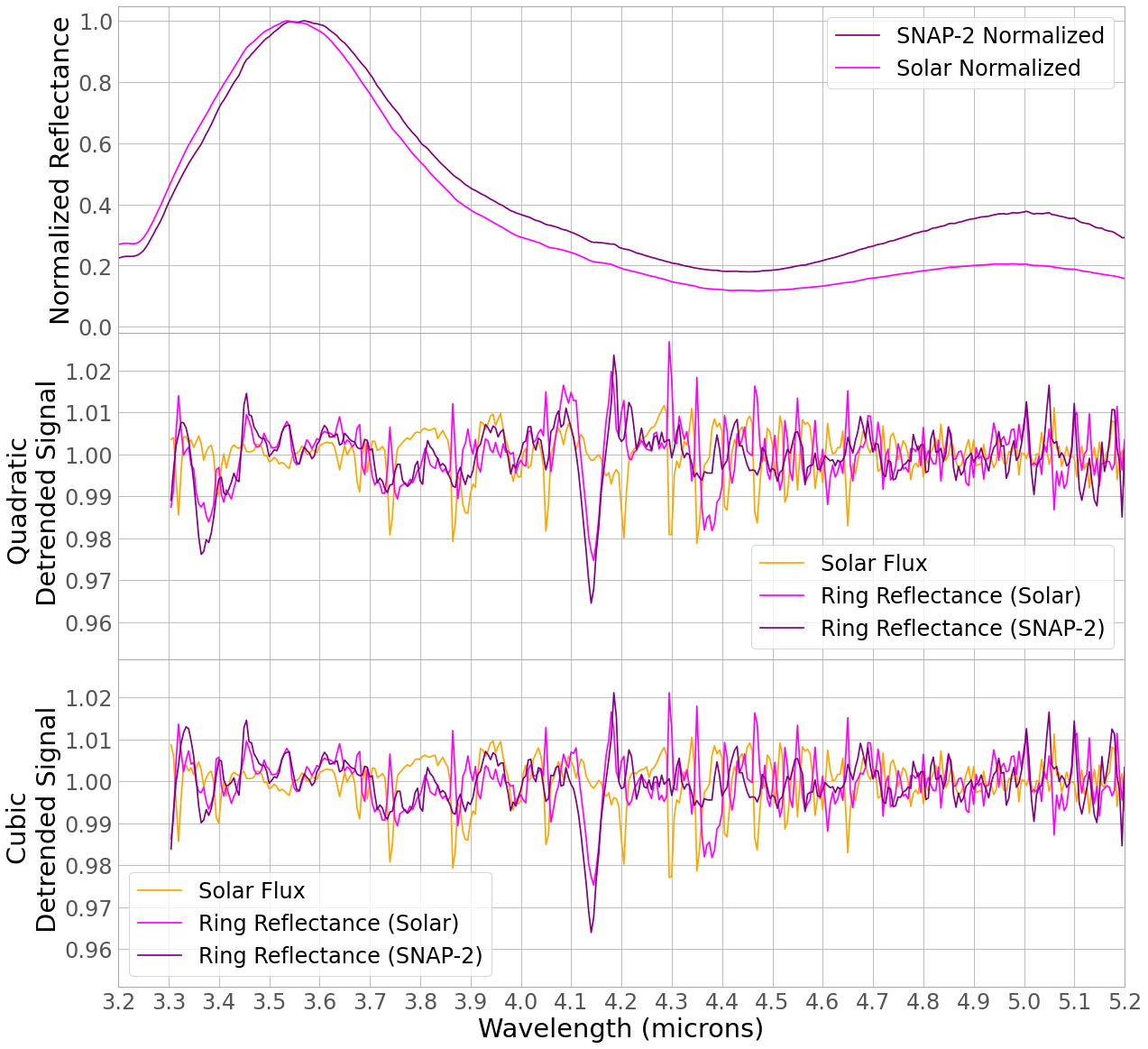}}
\caption{Normalized and de-trended near-infrared ring spectra. The top panel shows the normalized ring reflectance spectrum computed using a standard solar spectrum and the observed spectrum of a solar analog star SNAP-2. The differences between these two spectra arise because SNAP-2 is a slightly cooler star than the Sun. The lower panels show detrended versions of these spectra using rolling quadratic and cubic fits to the observed spectra (see text for details). There is a clear dip around 4.13 $\mu$m in all the detrended spectra due to the fundamental O-D absorption band. There is no evidence for a CO$_2$ band around 4.26 $\mu$m or a CO band around 4.7 $\mu$m. There may be a weak absorption band  around 3.4 $\mu$m that could potentially  be due to aliphatic hydrocarbons.}
\label{ringspec2}
\end{figure}

The overall shape of the observed spectrum, with broad peaks at 3.6 $\mu$m and 5.1 $\mu$m, is consistent with very pure water ice \cite{Mastrapa09, Clark12, Clark19}. This spectrum also shows a clear peak at 3.1 $\mu$m indicative of strongly crystalline water ice. This is consistent with Cassini-VIMS observations that showed this feature was much stronger in the ring specta than it was in most of the satellite spectra \cite{Filacchione12}. These data are of exceptionally high quality and clearly show two smaller peaks on either side of the 3.1 $\mu$m peak, which is consistent with the measured optical constants of highly crystalline water ice \cite{Mastrapa09, Clark12}.  

This spectrum covers a wavelength range where a number of more subtle spectral features have been detected in various Cassini-VIMS observations of icy objects in the Saturn system. More specifically, the fundamental O-D absorption band has been identified at 4.13 $\mu$m in very high signal-to-noise spectra of the B ring \cite{Clark19}. While the published ring spectra did not show any evidence for any absorption bands due to carbon-bearing compounds deeper than 1-3\% \cite{Clark08, Cuzzi09, Filacchione14, Clark19}, many of Saturn's moons exhibit a carbon dioxide band around 4.26 $\mu$m \cite{Clark08}, and Enceladus and Iapetus show aliphatic hydrocarbon signatures between 3.4 and 3.6 $\mu$m \cite{Brown06, Cruikshank14}. The new JWST spectrum of the A ring is of comparable quality to the highest signal-to-noise spectra obtained by Cassini-VIMS, and so it is worth examining whether these data contain any evidence for carbon-bearing compounds.

For this preliminary search for subtle spectral features, we convert the flux from the rings into estimates of the rings' reflectance using both the standard solar spectrum provided by STScI and the spectrum of the solar analog star SNAP-2 (2MASS J16194609+5534178) observed by JWST NIRSpec in the PRISM configuration as part of  program 1128 \cite{Gordon22}. SNAP-2 is a G3V star and so has a slightly lower effective temperature than the Sun. The reflectance spectrum computed using the SNAP-2 data therefore has a different overall slope than the one computed using the standard solar spectrum (see top panel of Figure~\ref{ringspec2}). Despite this, this spectrum is still useful for assessing whether any weak spectral features could be due to residual solar lines.

In order to better visualize and isolate any weak features in these spectra, we divided each reflectance spectrum at each wavelength channel by a rolling quadratic or cubic function that best fit the data within $\pm0.15$ $\mu$m of the selected wavelength channel, excluding any data below 3.3 $\mu$m (where the water-ice fresnel peaks complicate the fit) or between 4.1 and 4.2 $\mu$m (which contains the O-D band). This process removes the smooth variations in the spectral shape and normalizes the spectrum to a value of 1 at all wavelengths outside of sharp features. These normalized and detrended spectra are shown in the lower panels of Figure~\ref{ringspec2}, along with similarly normalized versions of the standard solar spectrum for reference. Note that the detrended spectra computed using the cubic and quadratic fits have similar shapes,  although the cubic versions have slightly less dispersion around unity, as is to be expected.

The most obvious feature in all the detrended spectra is the band centered at 4.13 $\mu$m, which corresponds to the O-D band identified by  \cite{Clark19} in Cassini-VIMS spectra of the rings. The depth of this band in the JWST data is between 2\% and 3\%, consistent with the prior analyses of the VIMS data \cite{Clark19}, which indicated that the D/H ratio for Saturn's rings is close to the terrestrial value. The JWST observations therefore confirm the existence and depth of this feature.

Outside of the O-D band, there are some differences between the detrended reflectance spectra computed using the standard solar spectrum and those computed using the SNAP-2 spectrum. Most notably, the spectra computed using the standard solar spectrum exhibit a series of narrow spikes that correspond to narrow dips in the predicted solar flux. Since these peaks are not seen in the spectra computed using SNAP-2,  they almost certainly represent residual solar lines. Similarly, the spectra computed using the solar reference spectrum show a 1.5\% dip at 4.37 $\mu$m that is not seen in the SNAP-2 normalized spectra. This feature is therefore  probably not a real spectral feature and may instead reflect an issue with the calibration pipeline that is still under active investigation (cf. Bockelee-Morvan et al. 2023). These issues imply that we should exercise due caution when interpreting $\sim1\%$ spectral features in these data sets.  Despite this, such weak features in these spectra can still provide useful information about potential non-water-ice components of the ring that may merit further investigation.
\nocite{BM23}

None of the detrended spectra show evidence of a carbon dioxide band around 4.26 $\mu$m or a carbon monoxide band around 4.7 $\mu$m. We may therefore conservatively conclude that both these bands must be weaker than 0.5\% in Saturn's A ring. These findings are consistent with prior analyses of the Cassini-VIMS observations \cite{Cuzzi09, Clark19}, and so confirm that oxygen-bearing carbon species are not currently detectable in Saturn's rings.

More interestingly, there are features between 3.3 $\mu$m and 4 $\mu$m in  all the detrended spectra. Two broad $\sim0.5\%$ dips are found around 3.75 $\mu$m and 3.97$\mu$m. The origins of these features  are not known at present.  A more sharp-edged band may also be present between 3.35 $\mu$m and 3.45 $\mu$m. This feature appears as a $\sim1\%$ dip in both spectra detrended using the quadratic fit, indicating that this feature is unlikely to be due to residual solar lines. While this feature is shallower in the spectra detrended using the cubic fit, its sharp edges are still visible even in these spectra, along with the narrow peak around 3.4 $\mu$m.   {Previous analyses of VIMS spectra found evidence of a weak potential absorption band around 3.42 $\mu$m band with a depth of $\sim3\%$ in the A ring \cite{Filacchione14} that was attributed to aliphatic hydrocarbons and could be consistent with this feature.} This structure also spans the same range of wavelengths as the aliphatic hydrocarbon absorption bands seen in the spectra of Enceladus and Iapetus  \cite{Brown06, Cruikshank14}, and has some similarities to the hydrocarbon bands recently observed by JWST on various  trans-neptunian objects  \cite{Emery23, BF23}. Detailed analysis and modeling of this feature could therefore potentially provide new constraints on the organic content of Saturn's rings.

\section{Mid-Infrared Observations of Saturn's Rings}
\label{miri}

\subsection{General description of observations}

JWST observed the rings at mid-infrared wavelengths as part of a MIRI observation sequence of both Saturn and its rings that is described in \citeA{Fletcher23}. The MIRI observations targeted at Saturn's rings were obtained on 2022-Nov-13 01:55:27 UT using the Medium-Resolution spectroscopy IFU mode, yielding spectra between 4.9 $\mu$m and 27.9 $\mu$m at a time when  the rings were at a phase angle of around 5.8$^\circ$ and the ring opening angle was 15$^\circ$. These observations used  the FASTR1 readout pattern and employed 5 groups per integration, 8 integrations per exposure and four dithers. This yielded a total exposure time for each channel of 521.708 seconds. The data from these observations are organized into four numbered channels covering overlapping wavelength ranges (Channel 1 = 4.9-7.65 $\mu$m, Channel 2 = 7.51-11.7 $\mu$m, Channel 3 = 11.55-17.98 $\mu$m, Channel 4 = 17.7-27.9 $\mu$m). Each of these channels consists of three sub-bands designated Short, Medium and Long, respectively, and have different spatial resolutions and fields of view (see Figure~\ref{miriim}).  Note that for this particular study we will not consider data from Channel 4 because the calibration of this channel is less reliable than the other channels. Also, a single dither using the same observation parameters (and so with a total exposure time of 130.427 seconds per sub-band) was obtained at 2022-Nov-13 03:17:19UT in a region of the sky 90'' north of Saturn that provides a measure of the background signal levels in the instrument. These cubes were all processed using pipeline version 1.11.4 with CDS context jwst\_1112.pmap. All observations were processed using the custom calibration pipeline described in \citeA{King23}.

\begin{figure}
\resizebox{\textwidth}{!}{\includegraphics{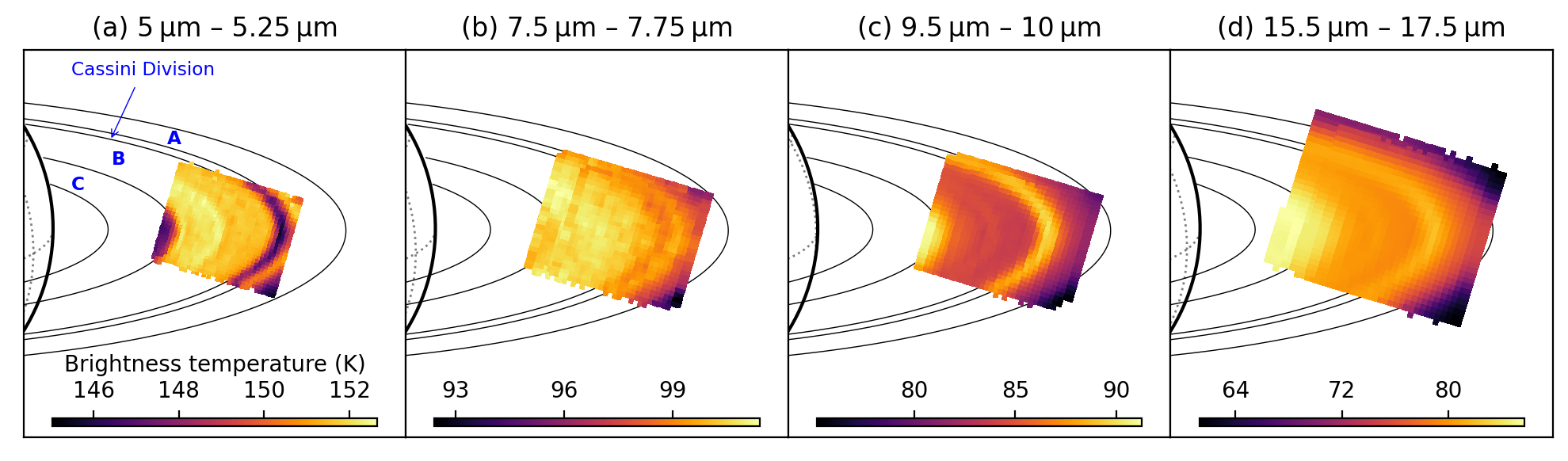}}
\caption{Mid-Infrared images of the rings derived from selected wavelength ranges. Note that at short wavelengths the A and B rings are brighter than the Cassini Division and C ring, while at long wavelengths the Cassini Division and C ring are brighter than the A and B rings. This contrast reversal arises because at short wavelengths the observed signal comes primarily from reflected sunlight, while at longer wavelengths the signal comes primarily from the rings' thermal emission.}
\label{miriim}
\end{figure}

\begin{figure}
\resizebox{\textwidth}{!}{\includegraphics{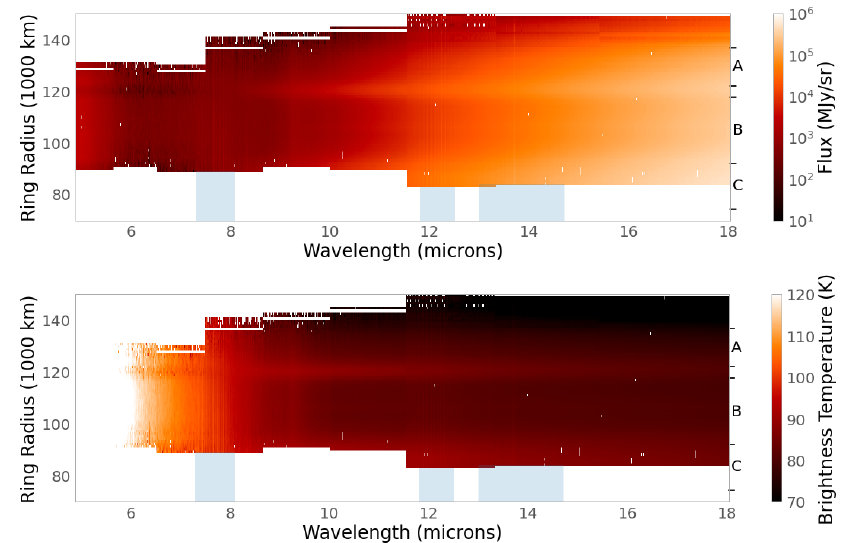}}
\caption{Maps of the ring's flux and brightness temperature as functions of ring radius and wavelength derived from the MIRI observations. The radial extent of the three main rings are shown at the far right of each plot (the Cassini Division is situated between the A and B rings). Note that the brightness temperature is more useful on the right-hand side of the figure, where thermal emission dominates over reflected sunlight. The brightness temperature rapidly increases at short wavelengths because of the increasing contribution of reflected sunlight. Narrow spectral features in the regions indicated by the shaded bands around 8 $\mu$m, 12 $\mu$m and 14$\mu$m are due to stray light from Saturn, but the bright line at 9.2 $\mu$m is due to the presence of highly crystalline water ice throughout the ring.}
\label{mirimap}
\end{figure}

Figure~\ref{miriim} shows images of the rings derived from a sample of the various wavelength regions covered by the MIRI observations. All these images are centered on the B ring, but also contain the Cassini Division and portions of the C and A rings. At the shortest observed wavelengths (around 5 $\mu$m) the A and B rings are both clearly brighter than the Cassini Division and C ring. This is consistent with the appearance of the rings at visible and near-infrared wavelengths \cite{Cuzzi09, Hedman13, Filacchione14, Cuzzi18, Ciarniello19}, and occurs because the dominant signal at these wavelengths is reflected sunlight. The Cassini Division and C ring are less reflective than the A and B rings both because they have lower optical depths and because the material in these regions is intrinsically darker \cite{Cuzzi09, Hedman13, Filacchione14}. However, at longer wavelengths the situation reverses, with the Cassini Division and C ring becoming brighter than the A and B rings around 10 $\mu$m. This happens because at these wavelengths the dominant signal comes from the ring's own thermal emission, and the lower albedos of the particles in the Cassini Division and C ring causes them to be warmer than the particles in the A and B rings  \cite{Spilker06, Altobelli08, Flandes10,  Filacchione14, Spilker18}. 

There are also comparatively subtle wavelength-dependent brightness variations within the B ring. Around 5 $\mu$m the inner B ring is slightly brighter than the outer B ring. This is notable because at wavelengths shorter than 1 $\mu$m the outer B ring is brighter at low phase angles \cite{Hedman13, Filacchione14}. This contrast reversal probably arises because the outer B ring has stronger water-ice absorption bands than the inner B ring \cite{Hedman13, Filacchione14}, which allows the contrast between the two regions to reverse at wavelengths longer than  3 $\mu$m, where the water ice is strongly absorbing {(see for example Observations X an II in Fig 4. of Filacchione et al. 2014).} At even longer wavelengths a faint bright band can be seen near the middle of the ring, which corresponds to regions of slightly elevated temperatures that were previously observed by Cassini-CIRS \cite{Spilker06, Spilker18}. 

{It is also worth noting that none of the rings show obvious brightness variations with longitude. This might at first be surprising because observations at multiple wavelengths indicate that both the A and B ring are composed of elongated canted structures called self-gravity wakes that generate azimuthal variations in the ring's opacity and brightness over a wide range of wavelengths \cite{Thompson81, LI84, Franklin87, Dones93, Dunn04, Nicholson05, French07, Ferrari09, Jerousek16}. In practice, however, such brightness variations will be subtle in these particular observations because the ring opening angle is sufficiently low that the B ring will be almost completely opaque at all observed azimuths. The A ring could exhibit substantial brightness variations in this viewing geometry, but these are difficult to discern in the MIRI data because substantial spans of A-ring longitudes are only observable at longer wavelengths, where the A ring is difficult to isolate from the brighter Cassini Division.}

{Another potential source of azimuthal brightness variations are changes in the temperature of the ring material as it emerges from Saturn's shadow and moves around the planet. While such temperature variations have been observed by Cassini-CIRS, it turns out these are quite small (1-2 K) over the range of longitudes observed by JWST (which correspond to local times around 18 hours)  when the lit side of the rings are observed  at low phase angles \cite{Leyrat08, Spilker18}. This is again consistent with the lack of obvious azimuthal brightness variations in the JWST data.}

{Given the lack of obvious azimuthal variations in the signal from the rings,} we transformed the calibrated MIRI cubes  into maps of the ring's brightness versus radius and wavelength. To accomplish this, we first subtracted instrumental backgrounds from each cube by subtracting the data from the offset cube, and then computed the average brightness at each wavelength for a series of 1000-km wide bins in the observed ring-plane radius. Figure~\ref{mirimap} show the resulting maps in two different formats. The top panel shows the measured flux from the rings in the standard units of MJy/sr.  In these units, the signal from each radius in the ring declines with increasing wavelength between 5 and 6.5 $\mu$m, which can be attributed to the decreasing reflectance of the rings over these wavelengths. However, beyond about 8 $\mu$m the signal strongly increases with wavelength, which is due to the steep increase in the ring's thermal emission over this wavelength range. Since the thermal emission flux spans several orders of magnitude, a clearer picture of the radial variations in the ring's spectra can be obtained by converting the observed flux into a ``Brightness Temperature", which is computed assuming the signal at each wavelength and location is entirely due to thermal emission from a single-temperature source that fills the appropriate region. At long wavelengths the ring's brightness temperature at each radius is approximately constant. However, at short wavelengths the brightness temperature increases with decreasing wavelength due to increasing importance of reflected sunlight.

The Cassini Division is clearly visible in both panels of Figure~\ref{mirimap} as a horizontal stripe around 120,000 km that transitions from dark to bright around 8 $\mu$m. The C ring can also be seen below 90,000 km that transitions from being dark to bright around the same wavelength. Note that the edge of the A ring around 140,000 km is rather diffuse due to limited spatial resolution of the instrument at long wavelengths.  At the same time, there are some features at particular wavelengths. There are multiple narrow bright features around 8 $\mu$m, 12 $\mu$m and 14 $\mu$m that turn out to be due to stray light from Saturn (see below). However, there is also a broader bright band around 9.2 $\mu$m that does appear to come from the rings themselves.

\begin{figure}
\resizebox{\textwidth}{!}{\includegraphics{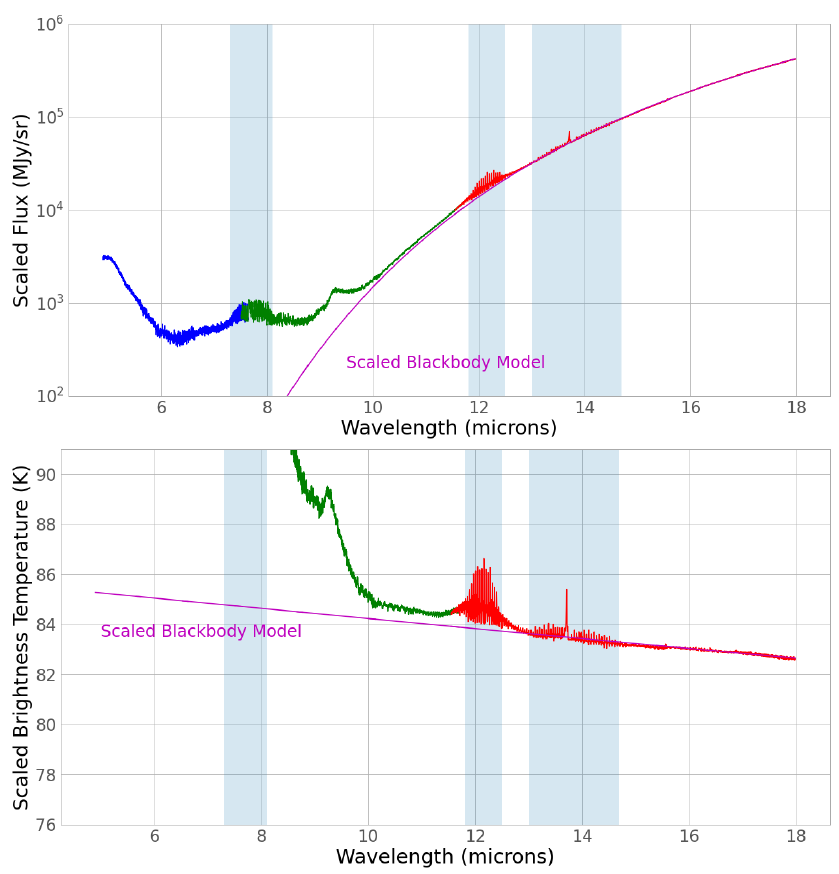}}
\caption{Average ring spectra of the middle B ring between 100,000 km and 110,000 km. Shaded bands indicate regions most obviously contaminated by emission lines in the stray light from Saturn. Note that the brightness in various sub-channels were scaled to fixed factors to match across each overlap region (see Table~\ref{scales}). These factors result in the longest wavelength channels being multipled by a factor of roughly 1.4 relative the shortest wavelength data. The bottom panel shows the scaled brightness spectra converted to brightness temperature, which {is not necessarily the physical temperature of the rings and} shows a steady decrease with increasing wavelengths even at wavelengths where thermal emission dominates. The magenta lines in both plots show an 86.3 K  blackbody spectrum multiplied by the scaling factor $\beta=0.663$ (see text). While this model reproduces the thermal component of the spectrum well, the scaling factor is lower than expected.}
\label{mirispec}
\end{figure}

\begin{table}
\caption{Scaling factors required for the B-ring spectra to match}
\label{scales}
\centerline{\begin{tabular}{|c|c|}\hline
Channel 1 Medium/Channel 1 Short & 1.081 \\
Channel 1 Long/Channel 1 Medium & 1.057 \\
Channel 2 Short/Channel 1 Long & 1.055 \\
Channel 2 Medium/Channel 2 Short & 1.087 \\
Channel 2 Long/Channel 2 Medium & 1.074 \\
Channel 3 Short/Channel 2 Long & 0.961 \\
Channel 3 Medium/Channel 3 Short & 1.025 \\
Channel 3 Long/Channel 3 Medium & 1.001 \\
\hline
\end{tabular}}
\end{table}

 \subsection{Mid-infrared spectra of Saturn's B ring}

The images and maps indicate that the Cassini Division is barely resolved, and the captured regions of the A and C rings do not have particularly uniform brightness. Interpreting the spectral trends in these regions would therefore require accounting for the wavelength-dependent point-spread function of MIRI/MRS, {as well as the azimuthal brightness asymmetries in the A ring \cite{Ferrari09}},  which is beyond the scope of this initial investigation. Hence for this particular analysis we will consider only the average spectrum of the region 100,000 and 110,000 km. This corresponds to the middle of the B ring, where the observed ring brightness is relatively uniform {in both radius and longitude} (note the variations in the brightness temperature across the B ring seen in Figure~\ref{miriim} are less than 2 K).  Figure~\ref{mirimap} shows the average ring flux and computed brightness temperature derived from the MIRI observations of this region. Note that the spectrum constructed from the individual channel sub-bands showed small offsets where they overlap with adjacent sub-bands. Since the ring signal fills multiple pixels and is rather strong at most of the sampled wavelengths, we interpret these offsets as due to differences in the calibration between the different channels, and so assume that multiplicative scaling factors are the appropriate way to bring these spectra into alignment (note that we cannot directly compare the signals measured by MIRI and NIRspec around 5 $\mu$m because the two instruments observed different parts of the rings.) Table~\ref{scales} shows the average brightness ratios between adjacent bands in these overlap regions {for the average B-ring spectrum}, which are all within 10\% of unity. We use these factors to re-normalize the various spectral channels relative to the shortest-wavelength sub-channel, which meant that the longest wavelength sub-channel is multiplied by a factor of 1.4. This is a relatively slight spectral slope over this wavelength range, and is not a significant issue for this particular analysis, which will focus primarily on the short-wavelength part of this spectrum. Also note that this correction is applied before computing the brightness temperature spectrum shown in the lower panel of Figure~\ref{mirispec}.

The average spectrum contains a number of interesting features. Again, the many narrow lines near 8 $\mu$m, 12 $\mu$m and 14 $\mu$m are due to stray light from Saturn, and correspond to strong emission lines from methane, ethane and acetylene, respectively \cite{Fletcher23}. These regions are shaded in Figures~\ref{mirispec}  because this contamination from the planet prevents these regions from providing reliable information about the composition of the rings. However, below 10 $\mu$m we can see two distinct peaks in the flux at 5 $\mu$m and 9.3 $\mu$m, along with a dip around 6.2 $\mu$m. These features are consistent with the expected reflectance spectra of crystalline water ice. However, before we can examine these features in detail we first need to consider the thermal component of these spectra.

At wavelengths longer than 10 $\mu$m, the flux increases smoothly over several orders of magnitude due to the ring's thermal emission. If this was a perfect blackbody spectrum, then the derived brightness temperature at long wavelengths would have a constant value corresponding to the temperature of the ring material. However, in practice  the brightness temperature above 10 $\mu$m shows an approximately linear decrease with increasing wavelength. This trend cannot be attributed solely to the scaling factors applied to the different sub-channels because this trend can be observed within individual sub-channels. The most straightforward explanation for this sort of trend is that the ring material either has a finite emissivity or does not completely fill the field of view due to the gaps between the ring particles. Both of these phenomena result in the observed thermal ring spectra being multiplied by a scaling factor $\beta$ that is less than 1 \cite{Spilker06, Spilker18}. In fact, if we fit the re-scaled data from sub-channel 3-Long (i.e. 15.41-17.98 $\mu$m) to a scaled blackbody spectrum, the best-fit model has a temperature of 86.3 K  and a $\beta$=0.663 (Note that  $\beta$ would be 0.478 if we did not re-scale the fluxes using the mean brightness ratios between bands). This scaled blackbody spectrum is a good match to the observed spectra in the regions free of {stray-light from Saturn} above 12 $\mu$m. 

However, while the shape of the spectrum is consistent with a scaled blackbody, the parameters derived from this fit are not entirely consistent with previous observations of the B-ring's thermal spectra obtained by the Cassini CIRS instrument. The temperature of 86.3 K is reasonably compatible with CIRS observations of the B ring taken at comparable phase and ring-opening angles, which are generally between 85 K and 90 K \cite{Spilker06, Altobelli08, Flandes10, Filacchione14, Spilker18}. The issue is that the $\beta$ value is well below the expected value for the B ring. The Cassini CIRS data indicates that the thermal scaling factor for the B ring can be predicted quite well by the ring's geometric filling factor based on optical depth measurements \cite{Spilker18}. For these particular observations of the B ring the filling factor is above 0.95 for the region between 100,000 and 110,000 km, which would imply $\beta>0.85$ \cite{Spilker18}, which is 35\% larger than the value needed to match the scaled MIRI spectra (This discrepancy grows to 80\% if we use $\beta=0.48$, which best fits the unscaled channel 3-Long data). The reasons for this discrepancy are still unclear,  but could include wavelength-dependent contamination of the spectra by either Saturn or blank sky due to MIRI's extended point-spread-function, both of which can potentially introduce spurious slopes and trends into the spectrum. Since these aspects of MIRI's performance are still under active investigation, we will not make further attempts to interpret the detailed shape of the thermal component of these spectra at this time. Instead, we will focus on the spectral features observed at wavelengths below 10 $\mu$m, whose shapes are relatively insensitive to uncertainties in the overall spectral slope.

\begin{figure}
\resizebox{\textwidth}{!}{\includegraphics{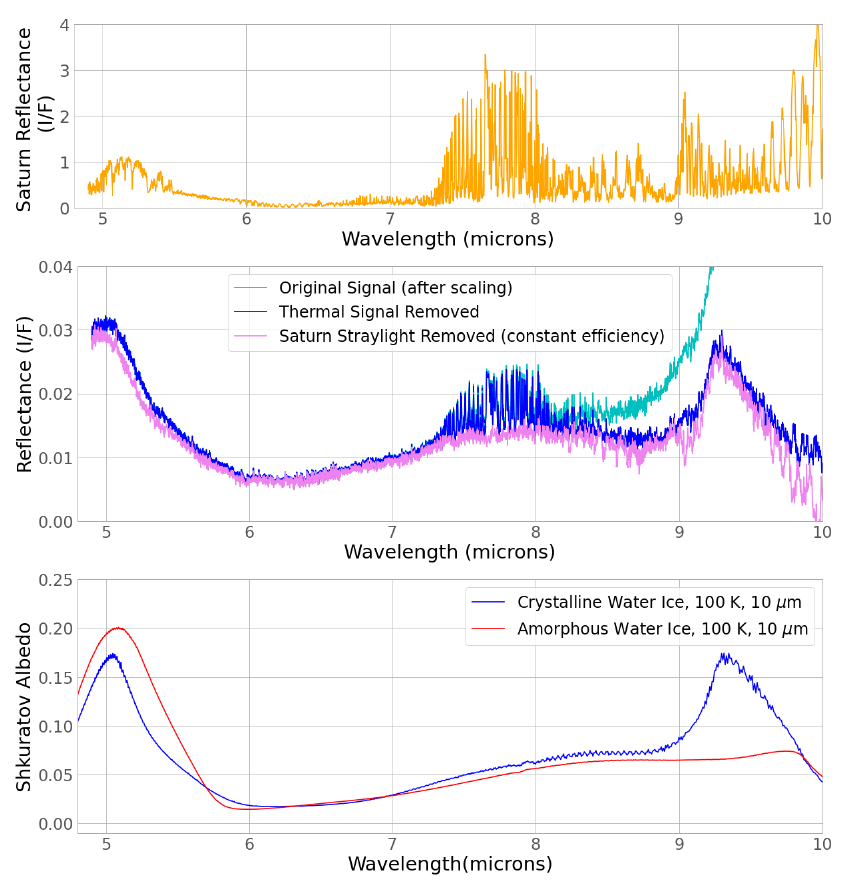}}
\caption{A closer look at the mid-infrared reflectance spectra of Saturn's rings. The top panel shows the average Saturn spectrum from the SATURN-15N observation, converted to equivalent $I/F$ for comparison with the rings. Note particularly the series of strong emission lines around 8 $\mu$m and beyond 9 $\mu$m. The middle panel shows three different versions of the average ring spectrum between 100,000 and 110,000 km. The cyan curve shows the brightness data after the scaling to remove offsets between subchannels. The blue curve shows the same curve after removing the  best-fit  average thermal signal with a temperature of 86.3 K and a $\beta=0.663$. The violet curve further has stray light from Saturn removed by subtracting the Saturn spectrum scaled by 0.0033. Note that this removed the lines around 8 $\mu$m, but introduces spectral features beyond 9 $\mu$m that might be spurious. The bottom panel shows the predicted spectra of crystalline and amorphous water ice regoliths with effective scattering lengths of 10 $\mu$m and temperatures of 100 K in this same wavelength range.}
\label{mirispec2}
\end{figure}

Figure~\ref{mirispec2} shows the average reflectance spectra of the rings between 100,000 and 110,000 km before and after removing the 86.3 K blackbody scaled by a factor of 0.663. The conversion from flux density to reflectance again uses the standard solar spectrum available from STScI (https://archive.stsci.edu/hlsps/reference-atlases/cdbs/grid/solsys/). Note that removing the thermal signal only strongly affects the signal at wavelengths longer than  8 $\mu$m. In addition, it is reasonable to assume $\beta$ is independent of wavelength because the generally low reflectance of the rings at these wavelengths means that the emissivity should be close to unity throughout  this wavelength range \cite{Morishima12}. Finally, we can note that around 5 $\mu$m, the peak reflectance is around 0.03, which is consistent with the near-infrared spectrum shown in Figure~\ref{ringspec}, especially if we realize that the B ring is slightly brighter than the A ring at near-infrared wavelengths when viewed at low phase angles \cite{Filacchione12, Hedman13, Filacchione14}.

Stray light from Saturn is clearly present in these spectra around 8 $\mu$m (corresponding to stratospheric methane emission), but contamination from the planet also may be having more subtle effects on other parts of this spectrum. For comparison, Figure~\ref{mirispec2}  also shows a Saturn spectrum derived from the average brightness in the cubes targeted at low Saturn latitude (designated SATURN-15N), converted to reflectance to facilitate comparisons with the ring spectrum. While the lines around 8 $\mu$m are the obvious features in the Saturn spectrum, there are also features in the Saturn spectrum that could be influencing the ring spectrum around 5 $\mu$m and 10 $\mu$m. In order to estimate how much stray-light from Saturn could be influencing the ring spectrum, we estimate the likely strength of the Saturn signal in the ring spectrum {by plotting the ring's brightness versus Saturn's brightness in the wavelength range between 7.25 and 8.15 $\mu$m and fitting these data to a linear trend.} This fit indicated that the lines in the ring spectrum are about 0.0033 the intensity of the lines in the Saturn spectrum. We therefore subtracted a version of the Saturn spectrum scaled by this factor from the ring spectrum to produce the violet curve in Figure~\ref{mirispec2}. This correction changes the shape of the 5 $\mu$m peak slightly, and also introduces narrow features into the ring spectrum between 8.5 $\mu$m and 10 $\mu$m. The latter implies that assuming a fixed scaling factor is overcorrecting the ring spectrum beyond 8.5 $\mu$m. This is likely due to a combination of variations in Saturn's spectrum across its disk \cite{Fletcher23}  and wavelength-dependent variations in the MIRI point-spread function. The stray-light signals from Saturn therefore vary with wavelength in  rather complex ways that will need to be addressed in future work. Fortunately the overall shape of the peak at 9.3 $\mu$m is insensitive to details of this subtraction.

Comparing all  these spectra with the predicted spectra of amorphous and crystalline water ice computed using the \citeA{Mastrapa09} optical constants and the \citeA{Shkuratov99} scattering formulas, we can see that the overall shape of the observed spectra matches those expected for crystalline ice extremely well. Most dramatically, the peak in the observed spectrum at 9.3 $\mu$m is only found in the crystalline ice spectrum, and is absent in amorphous ice, providing independent confirmation that the ice in Saturn's rings is strongly crystalline. The shape of the 5 $\mu$m peak (after correcting for Saturn shine) is also closer to that of the crystalline model than the amorphous model.  Differences between the crystalline ice model and the observed ring spectrum in this range can mostly be attributed to issues with removing the thermal signal and the straylight from Saturn, particularly at  longer wavelengths.  Since these spectra match crystalline water ice so well, they can serve as a useful baseline for interpreting mid-infrared spectra of other bodies in the outer solar system with more complex compositions.

\section{Summary of Key Findings}
\label{summary}

The key findings of this initial analysis of the JWST spectra of Saturn's rings and small moons are as follows:

\begin{itemize}
\item NIRSpec could clearly detect and  measure spectral features of Saturnian moons with radii between 60 km and 2 km {(or angular radii between  8.5 and 0.3 milliarcseconds)}.

\item The near-infrared spectra of Epimetheus, Pandora and Telesto are dominated by water ice features, consistent with prior observations, and the shapes of the 1.5 $\mu$m and 2.0 $\mu$m bands, as well as the lack of clear 3.1 $\mu$m peaks, suggest  that either the surfaces of these moons are composed of a mixture of crystalline and amorphous water ice, or the surface includes significant amounts of sub-micron grains {and/or contaminants}. Overall, Telesto appears to have a  more crystalline spectrum than Epimetheus does.

\item The near-infrared spectrum of Pallene is also dominated by water ice features, and its water ice bands are similar to those previously observed on the nearby and comparably small moon Methone. The shape of Pallene's  2 $\mu$m band suggests that its surface ice {could be} more amorphous than the other small moons observed by JWST, which may happen because Pallene's surface is exposed to relatively high doses of high-energy radiation \cite{Hedman20}.

\item The near-infrared spectrum of Saturn's A ring is dominated by features consistent with highly crystalline water ice, including a prominent set of {Fresnel} peaks around 3.1 $\mu$m.

\item The JWST ring spectrum confirms the existence and depth of the 4.13 $\mu$m band due to O-D absorption, and therefore supports previous work suggesting that the rings have a D/H ratio close to terrestrial values \cite{Clark19}.

\item {The JWST near-infrared ring spectrum does not show any clear spectral features at 4.26 $\mu$m due to carbon dioxide or at 4.7 $\mu$m due to carbon monoxide. However, there may be a weak feature around 3.4 $\mu$m that could be due to aliphatic hydrocarbons.}

\item The mid-infrared ring observations show a clear contrast reversal, with the A and B rings being brighter than the Cassini Division and C ring at shorter wavelengths due to their higher reflectance, while the Cassini Division and C ring are brighter at longer wavelengths because of their higher temperatures.

\item The rings have a reflectance peak at 9.3 $\mu$m due to highly crystalline water ice, and the overall shape of the B-ring spectrum between 5 and 10 $\mu$m is consistent with very pure, highly crystalline water ice. More work is needed to understand the stray light from Saturn and overall radiometric calibration of MIRI.

\end{itemize}

\section*{Open Research Section}
Level-3 calibrated data from the standard pipeline are available directly from the MAST archive {\tt https://mast.stsci.edu}. The JWST calibration pipeline is described in \citeA{Bushouse23}, and the custom pipeline and initial data processing code is described in \citeA{King23}.  The Jupyter notebooks used to produce the spectra shown in this study, along with relevant input files and csv files of the flux spectra for the rings and small moons, are available at {\tt https://github.com/JWSTGiantPlanets/SaturnRingsMoons} and via \citeA{Hedman24}. 

\section*{Acknowledgments}
Fletcher, King, and Roman were supported by a European Research Council Consolidator Grant (under the European Union's Horizon 2020 research and innovation programme, Grant 723890) at the University of Leicester. Harkett was supported by an STFC studentship; Hammel and Milam acknowledge support from NASA JWST Interdisciplinary Scientist Grant 21-SMDSS21-0013. We wish to express our gratitude to the JWST support team for their patience and perseverance as we designed these observations—in particular Beth Perriello, Bryan Holler, Stephen Birkman, Misty Cracraft, Tony Roman, and John Stansberry for their aid in setting up the observations in APT, and David Law for his tireless support as we developed codes to interpret MIRI/MRS data. This research used the ALICE High Performance Computing Facility at the University of Leicester. This work is based on observations
made with the NASA/ESA/CSA JWST. The data were obtained from the Mikulski Archive for Space Telescopes at the Space Telescope Science Institute, which is operated by the Association of Universities for Research in Astronomy, Inc., under NASA contract NAS 5-03127 for JWST. These observations are associated with program 1247 (PI: Fletcher). We also wish the thank Mauro Ciarniello and Gianrico Filacchione for their helpful comments on an earlier version of this manuscript.

\bibliography{JWSTsat}

\end{document}